\documentclass[manuscript=article]{achemso}

\usepackage[version=3]{mhchem} 
\usepackage{graphicx}
\usepackage{amsmath}
\usepackage{amssymb}
\usepackage{xcolor}
\usepackage{amsmath}
\usepackage{physics}
\usepackage{lineno,hyperref}
\usepackage{multirow}
\usepackage{tabularx}
\usepackage{longtable}      
\usepackage[dvipsnames]{xcolor} 
\usepackage[dvipsnames]{xcolor}
\usepackage[utf8]{inputenc}
\usepackage[english]{babel}
\usepackage[T1]{fontenc}
\usepackage{amssymb}
\usepackage{multirow}
\usepackage{longtable}
\usepackage{graphicx}
\graphicspath{ {./images/} }
\usepackage{caption}
\usepackage{subcaption}
\usepackage{gensymb}
\usepackage{pdfpages}
\usepackage{siunitx}

\author{Syed Mustafa Shah}
\altaffiliation{These authors contributed equally to this work.}
\affiliation[a]
{Department of Chemical Engineering, University of Illinois Chicago, Chicago, IL 60608, USA}

\author{Musawenkosi K. Ncube}
\altaffiliation{These authors contributed equally to this work.}
\affiliation[a]
{Department of Chemical Engineering, University of Illinois Chicago, Chicago, IL 60608, USA}

\author{Mohammed Lemaalem}
\altaffiliation{These authors contributed equally to this work.}
\affiliation[a]
{Department of Chemical Engineering, University of Illinois Chicago, Chicago, IL 60608, USA}

\author{Selva Chandrasekaran Selvaraj}
\affiliation[a]
{Department of Chemical Engineering, University of Illinois Chicago, Chicago, IL 60608, USA}

\author{Naveen K. Dandu}
\affiliation[a]
{Department of Chemical Engineering, University of Illinois Chicago, Chicago, IL 60608, USA}

\author{Alireza Kondori}
\affiliation[b]
{Department of Chemical and Biological Engineering, Illinois Institute of Technology, Chicago, IL 60616, USA}

\author{Gayoon Kim}
\affiliation[b]
{Department of Chemical and Biological Engineering, Illinois Institute of Technology, Chicago, IL 60616, USA}

\author{Adel Azaribeni}
\affiliation[b]
{Department of Chemical and Biological Engineering, Illinois Institute of Technology, Chicago, IL 60616, USA}

\author{Mohammad Asadi}
\affiliation[b]
{Department of Chemical and Biological Engineering, Illinois Institute of Technology, Chicago, IL 60616, USA}

\author{Anh T. Ngo}
\email{anhngo@uic.edu}
\affiliation[a]
{Department of Chemical Engineering, University of Illinois Chicago, Chicago, IL 60608, USA}
\alsoaffiliation[c]
{Materials Science Division, Argonne National Laboratory, Lemont, IL 60439, USA}

\author{Larry A. Curtiss}
\email{curtiss@anl.gov}
\affiliation[c]
{Materials Science Division, Argonne National Laboratory, Lemont, IL 60439, USA}

\affiliation[a]{Department of Chemical Engineering, University of Illinois Chicago, Chicago, IL 60608, USA}
\affiliation[b]{Department of Chemical and Biological Engineering, Illinois Institute of Technology, Chicago, IL 60616, USA}
\affiliation[c]{Materials Science Division, Argonne National Laboratory, Lemont, IL 60439, USA}
\title{Atomic-Scale Mechanisms of Li-Ion Transport Mediated by Li$_{10}$GeP$_{2}$S$_{12}$ in Composite Solid Polyethylene Oxide Electrolytes}
\begin{document}
\section*{Abstract}
Polymer electrolytes incorporating Li$_{10}$GeP$_{2}$S$_{12}$ (LGPS) nanoparticles are promising for solid-state lithium batteries due to their potential for enhanced ionic conductivity; yet, the atomistic mechanisms driving this enhancement remain debated. Here, we systematically investigate the relationship between LGPS nanoparticle loading, polyethylene oxide microstructure, and Li-ion transport using a combination of molecular dynamics (MD) simulations, experimental ionic conductivity measurements, and density functional theory (DFT) calculations. MD simulations and experiments reveal good agreement on ionic conductivity as a function of LGPS concentrations of up to 10 weight \% (x\%), exhibiting a volcano-like curve with ionic conductivity increasing fivefold from the low concentrations and can be accounted for by a  classical transport mechanism governed by polymer segmental dynamics and interface effects. However, at more than 10\% LGPS, experiments show further conductivity enhancement that cannot be accounted for by MD simulations, indicating a shift to another  transport mechanism. DFT calculations elucidate that, at the polymer|LGPS interface, Li-ion migration proceeds via vacancy-driven hopping, with barriers sensitive to local atomic composition-low-barrier pathways are possible when S atoms dominantly occupy the sites on the interface to facilitate  Li-hopping, while pathways involving Ge act as obstacles to Li transport. These results establish that optimized interfacial chemistry and electrolyte structure enable efficient, barrier-lowering migration channels that are distinct from bulk polymer or ceramic behavior. Our approach reconciles experiments with classical simulations at low LGPS concentrations and quantum chemical interface calculations, highlighting design criteria for maximizing the performance of these types of solid composite polymer electrolytes and guiding the development of advanced lithium batteries.
\begin{tocentry}
\centering
    \includegraphics[width=0.9\textwidth]{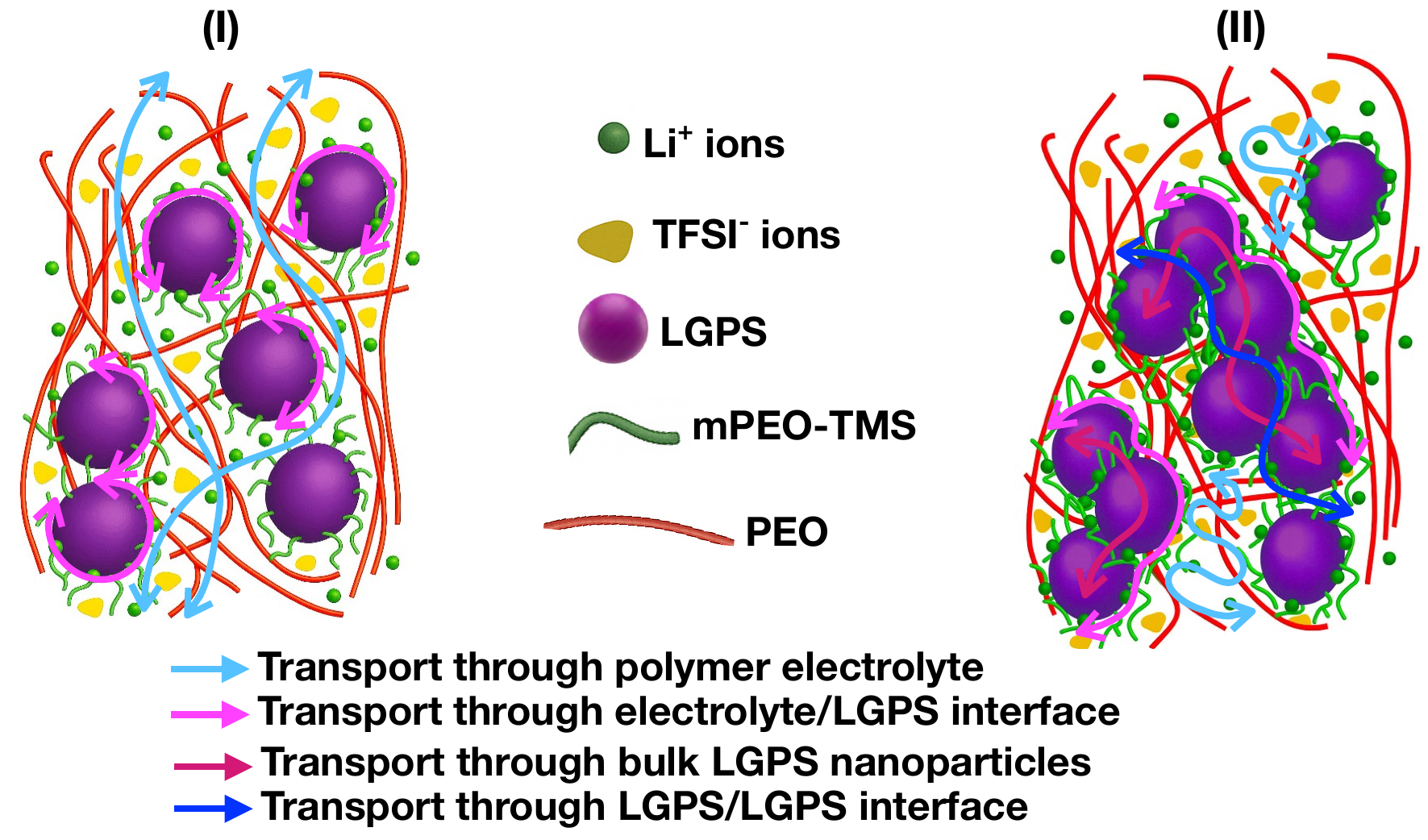}

\end{tocentry}
\section*{Introduction}
All-solid-state batteries (ASSBs) are among the most promising next-generation energy storage systems, offering high energy density, enhanced safety, and long cycle life \cite{kondori2023room, zhang2025fiber, zhou2025molecular, lemaalem2023graphyne, lemaalem2023tunable}. Within ASSBs technology, solid composite polymer electrolytes (CPEs), which incorporate inorganic lithium-ion conductors into polymer matrices, have attracted significant attention for their ability to combine the mechanical flexibility of polymers with the high ionic conductivity of inorganic phases. Among the various inorganic fillers, Li$_{10}$GeP$_{2}$S$_{12}$ (LGPS) stands out as one of the most conductive sulfide electrolytes, exhibiting room temperature ionic conductivities exceeding \(10^{-2}\,\text{S\,cm}^{-1}\)~\cite{Kato2016, Seino2014, Kuhn2013, Yu2021, Schweiger2022}.

A wide range of experimental and theoretical studies have explored the ion transport mechanisms in solid polymer composite electrolytes containing nanoparticles. The addition of inorganic fillers such as LGPS can disrupt polymer crystallinity, enhance the amorphous content, and increase segmental motion, thereby facilitating cation migration~\cite{Zhou2018, Bhattacharyya2020, zhang2019sulfide}. At moderate loadings, nanoparticles may create percolative ion-conduction networks and provide high-dielectric environments that increase the concentration of free lithium ions~\cite{Dawson2022, Cui2023, zhang2019sulfide}. Several reports also highlight that the interfacial region between the polymer and nanoparticles is critical: ion transport can be improved through tailored nanoparticle-polymer interactions, as well as functionalization of the filler to optimize compatibility and interfacial ionic conduction pathways~\cite{zhang2019sulfide, Hori2022, pan2020flexible}. However, excessive nanoparticle content often leads to agglomeration, restricted polymer dynamics, and diminished conductivity, as the nanoparticles can immobilize nearby polymer chains and hinder segmental motion~\cite{Zhou2018, Bhattacharyya2020}. Morphological control of the filler, such as using nanowires or aligned frameworks rather than particles, has also been shown to provide continuous and highly conductive pathways for lithium ions~\cite{zhang2019sulfide, Dawson2022}. \\

Nevertheless, the precise atomistic mechanisms underpinning ion transport in nanoparticle-polymer composite electrolytes, including the role of the interface and optimal filler concentration, remain insufficiently understood. In this work, we employ a combined atomistic molecular dynamics (MD) and density functional theory (DFT) framework to elucidate the multiscale mechanisms governing ion transport in a CPE comprising LGPS nanoparticles, polyethylene oxide (PEO), methoxy-poly(ethylene oxide)-trimethoxysilane (mPEO-TMS), and Lithiumbis(trifluoromethanesulfonyl)imide (LiTFSI) used in a solid-state Li-air battery \cite{kondori2023room}. The mPEO-TMS is used to functionalize the LGPS nanoparticles to stabilize them and enhance interfacial conduction.  This CPE is similar to another one developed for use in a Li-ion battery \cite{pan2020flexible}. In that study, the LGPS functionalization was performed using polyethylene glycol (PEG) and (3-chloropropyl)trimethoxysilane (CTMS). 

The MD simulations on the mPEO-TMS system capture the influence of nanoparticle loading on polymer structure, local coordination, and dynamic parameters such as ionic conductivity and cation transference number. Complementary DFT calculations probe Li-ion migration at the mPEO-TMS|LGPS interface, providing insights into interfacial bonding and migration barriers. Together, these approaches establish a mechanistic link at low LGPS concentrations between nanoscale interfacial chemistry, polymer segmental motion, and macroscopic ion transport, guiding the rational design of next-generation solid-state polymer electrolytes with optimized filler loading and interface functionality.

\section{Computational framework and ionic conductivity measurements}
\subsection{MD Simulation Details}
Atomistic molecular dynamics (MD) simulations were performed to model a composite solid polymer electrolyte comprising Li$_{10}$GeP$_{2}$S$_{12}$ (LGPS) nanoparticles, poly(ethylene oxide) (PEO), methoxy-poly(ethylene oxide)-trimethoxysilane (mPEO-TMS), and LiTFSI salt (Figure~\ref{fig:MD1}), based on the experimental composition presented in the supplementary information (Tables S1 and S2). In this study, we investigated a broader range of LGPS loadings from 0 to 40 weight \% (x\%) (relative to the total weight of all electrolyte components, except the salt) to assess the effect of filler content on ion transport. For computational feasibility, the LGPS particle size of 1.2\,nm was used while preserving the experimental mass ratios among all components. We note that the experimental LGPS crystallite size is 17 nm, but the actual nanoparticle size is not known \cite{kondori2023room}. When rescaling the LGPS nanoparticle size, the molecular weight of the PEO chains was also proportionally modified to match the LGPS-rescaled molecular weight, ensuring consistent polymer--filler mass relationships. The other small-molecule species, namely mPEO-TMS and LiTFSI, retained their original molecular models; only their quantities were adjusted to ensure that the experimental weight composition was strictly maintained. The simulation cell contained a set number of LGPS nanoparticles dispersed in polymer-salt matrix, as presented in Table \ref{Table1}. All simulations were carried out using LAMMPS~\cite{thompson2022lammps} with the OPLS-AA force field for organic species \cite{doherty2017revisiting} and the UFF potential for LGPS and TMS \cite{rappe1992uff}, more computational details are discussed in the Supplementary Information.
\begin{figure*}[!ht]
    \centering
    \includegraphics[width=1\textwidth,]{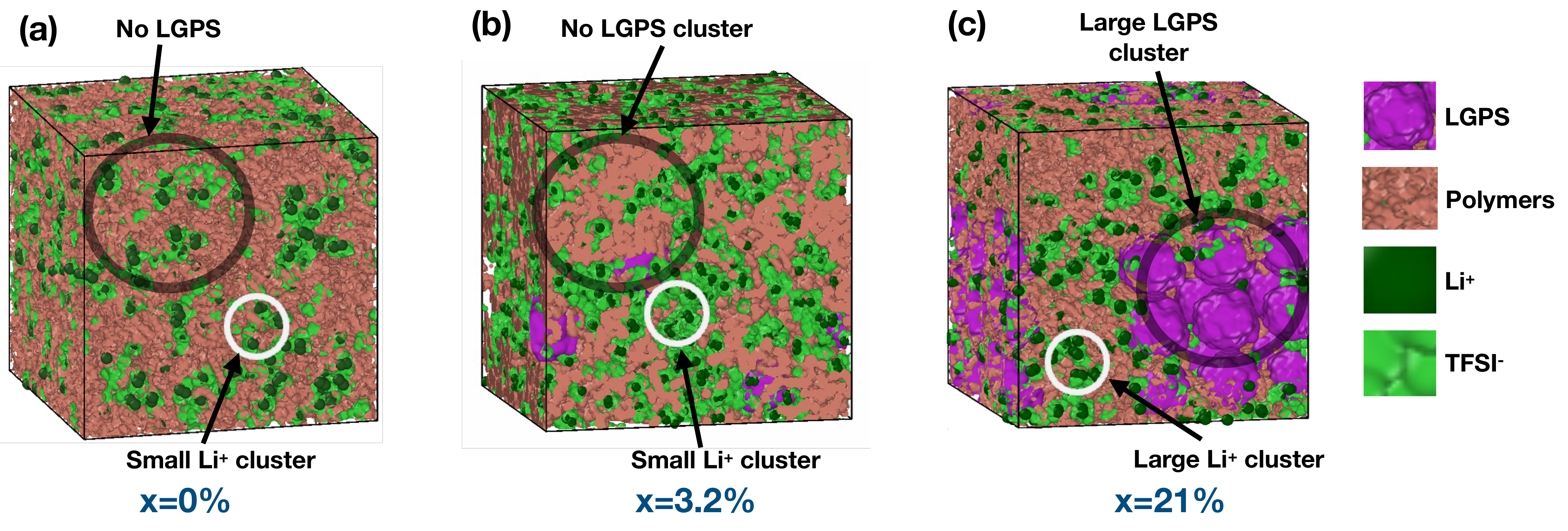}
    \caption{Visualization of the MD simulations of the composite polymer electrolyte (CPE) for different LGPS weight ratio (x\%): (a) x=0\%, (b) x=3.2\% and (c) x=21\%. Note that "Li$^{+}$ cluster" noted in the figures includes also TFSI$^{-}$ that are interacting with the Li$^{+}$ cations.}
    \label{fig:MD1}
\end{figure*}

\begin{table}[h!]
\centering
\caption{Number of particles in each simulation system.}
\label{tab:particles_simple}
\begin{tabular}{|c|c|c|c|c|c|c|c|c|c|}
\hline
\textbf{Element} & \textbf{0\%} & \textbf{1.3\%}  & \textbf{2\%} & \textbf{3.20\%} & \textbf{9\%} & \textbf{17\%} & \textbf{21\%} & \textbf{30\%} & \textbf{40\%} \\
\hline
\textbf {LGPS}     & 0    & 1    & 2    & 3   & 8   & 18  & 24  & 39 & 60  \\
\textbf {mPEO-TMS} & 1080 & 1080 & 1080 & 1080 & 1080 & 1080 & 1080 & 1080 & 1080  \\
\textbf {LiTFSI}   & 2400 & 2400 & 2400 & 2400 & 2400 & 2400 & 2400 & 2400 & 2400 \\
\textbf {PEO}      & 128  & 128  & 128  & 128  & 128  & 128  & 128 & 128 & 128 \\
\hline
\end{tabular}
\label{Table1}
\end{table}
\subsection{DFT calculations details}
We carried out spin-polarized DFT calculations to optimize the structures as well as to determine the Li cation migration pathway. The initial LGPS bulk structure was sourced from the Materials Project and cut using Crystal Maker to a 1.2~nm particle comprising 400 atoms~\cite{Jain2013}. The DFT calculations implemented the Projector-Augmented Wave (PAW) potentials for the elements as supplied by the Vienna Ab Initio Simulation Package (VASP)~\cite{Kresse1996b, Kresse1999}. We used the Generalized Gradient Approximation (GGA), based on the Perdew–Burke–Ernzerhof approach, to account for the exchange-correlation energies~\cite{Perdew1996}. The kinetic energy cutoff was set to 600~eV to enhance accuracy, and the Brillouin zone was sampled at the gamma point with the k-mesh set based on the respective lattice parameters. The energy convergence thresholds for the electronic self-consistent loop and ionic relaxation loop were set to 0.01~meV/\AA{} and 0.1~meV/\AA{}, respectively. All the structures were visualized using VESTA~\cite{Momma2011}.
\begin{figure*}[!ht]
    \centering
    \includegraphics[width=1\textwidth,]{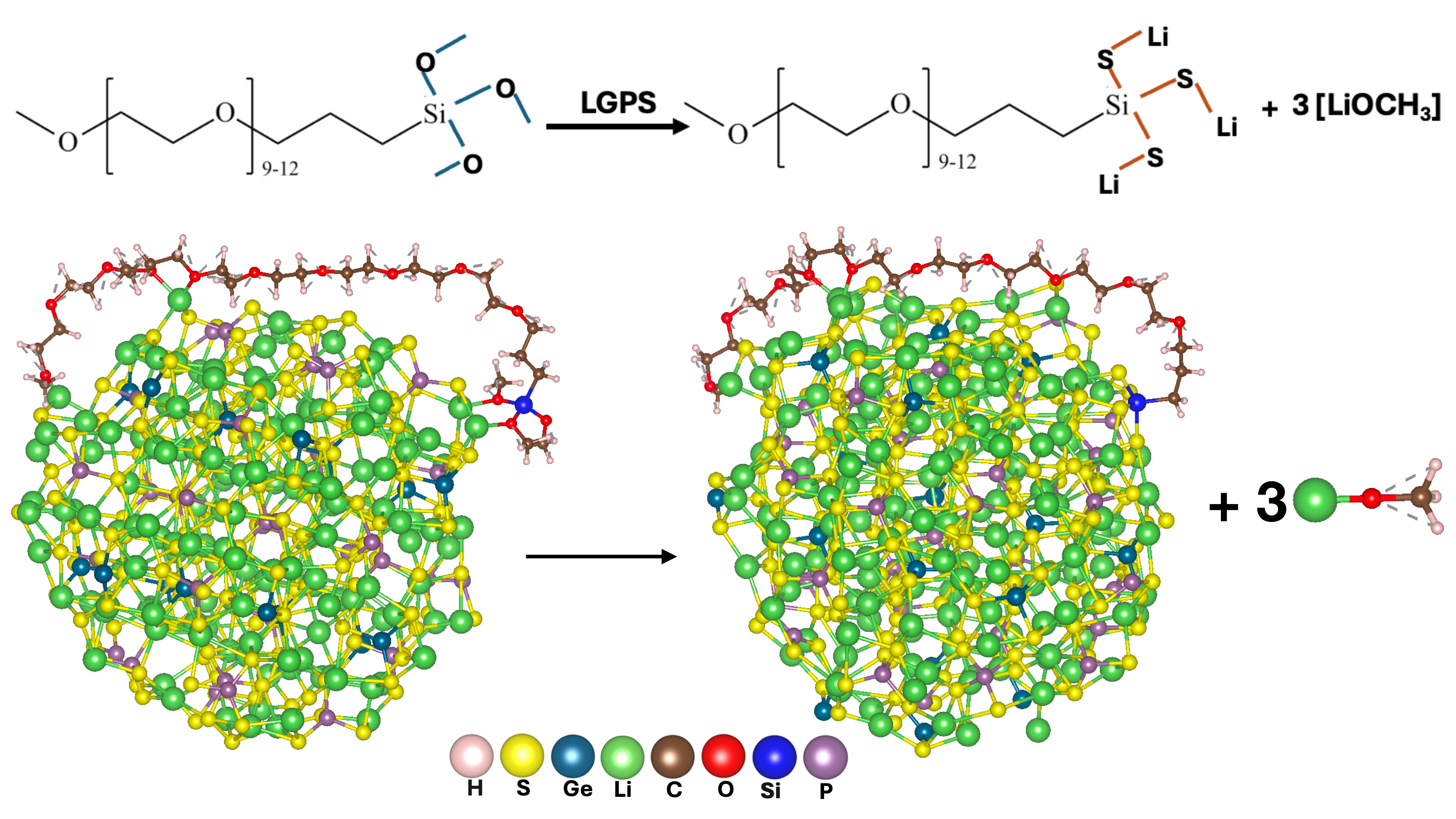}
    \caption{Reaction of 3-[methoxy(polyethyleneoxy)$_{6-9}$ propyl] trimethoxysilane (mPEO-TMS) with Li$_{10}$GeP$_2$S$_{12}$ to form a mPEO-TMS|LGPS interface coupled together by a SiS$_3$ bond and some Li-O bonds from wrapping around the nanoparticle, while releasing 3 molecules of LiOCH$_3$. In the top reaction, the elements in bold are part of the LGPS particle.}
    \label{fig:FigDFT1}
\end{figure*}
The surface of the LGPS nanoparticles was modified using the silane coupling agent mPEO-TMS. In Ref.~\cite{kondori2023room} evidence was presented  that the Si atoms in mPEO-TMS form  bonds with the S atoms in LGPS, thereby stabilizing the nanoparticle to prevent decomposition at the Li anode or cathode ~\cite{kondori2023room}. As illustrated in Figure~\ref{fig:FigDFT1}, this surface reaction results in the formation of SiS$_3$ bonds and the release of three LiOCH$_3$ molecules. The mPEO-TMS/LGPS structure based on a 1.2 nm particle, with its geometry optimized, was used to calculate energy barriers for the Li cation transport along the mPEO-TMS|LGPS interface. We used the nudged elastic band (NEB) method described by Sheppard \textit{et al.}~\cite{Sheppard2008} to determine the minimum energy paths. The NEB method uses a string of images (geometric configurations with varying atomic positions) between the initial and final states to describe the migration pathway. These images are connected by spring forces, which ensure equal spacing along the path. They are minimized using DFT and compared with the initial state energy to determine the energy required for the ion to move along the interface.

\subsection{Ionic conductivity measurements}

The ionic conductivity of the CPE was measured by electrochemical impedance spectroscopy (EIS, AC impedance) using a symmetric SS\(|\)CPE\(|\)SS cell, following the procedure described in Ref.~\cite{kondori2023room}. Measurements were performed at room temperature over a frequency range of 100~kHz to 0.1~Hz with an AC amplitude of 5 mV using a BioLogic SP150 potentiostat, which provides an adequate signal-to-noise ratio while maintaining a linear current response. The ionic conductivity \(\sigma\) of the CPE was measured as: $\sigma = \frac{L}{A\,R_{\mathrm{s}}}$, where \(L\) (cm) and \(A\) (cm\(^2\)) are the thickness and the electrode–electrolyte contact area of the CPE, respectively, and \(R_{\mathrm{s}}\) (\(\Omega\)) is the bulk resistance. Ionic conductivity from these measurements are shown in Figure~\ref{fig:MD2} for the solid electrolyte composed of poly(ethylene oxide) (PEO), 3-[methoxy(polyethyleneoxy)\(_{6-9}\)]propyltrimethoxysilane (mPEO), and Li\(_{10}\)GeP\(_2\)S\(_{12}\) (LGPS) as a function of LGPS content. The LGPS mass was varied from 0~mg to 400~mg, while the mPEO mass was kept constant at 1~g. In all samples, 0.5 g of PEO with a molecular weight of 1M was used. These measurements were also carried out in an SS\(|\)CPE\(|\)SS cell at room temperature.
\section{Results and discussions}
\subsection{Transport properties from MD simulation}
We analyzed transport in electrolytes using Onsager transport coefficients. These coefficients, denoted as $L^{ij}$, provide a more direct physical interpretation of ion correlations and can be computed directly from molecular simulations using Green-Kubo relations. Our primary goal is to investigate the electrolyte's dynamic properties, including ionic conductivity and cation transference number, which directly impact battery electrochemical performance.
\begin{figure*}[!ht]
    \centering
    \includegraphics[width=0.5\textwidth,]{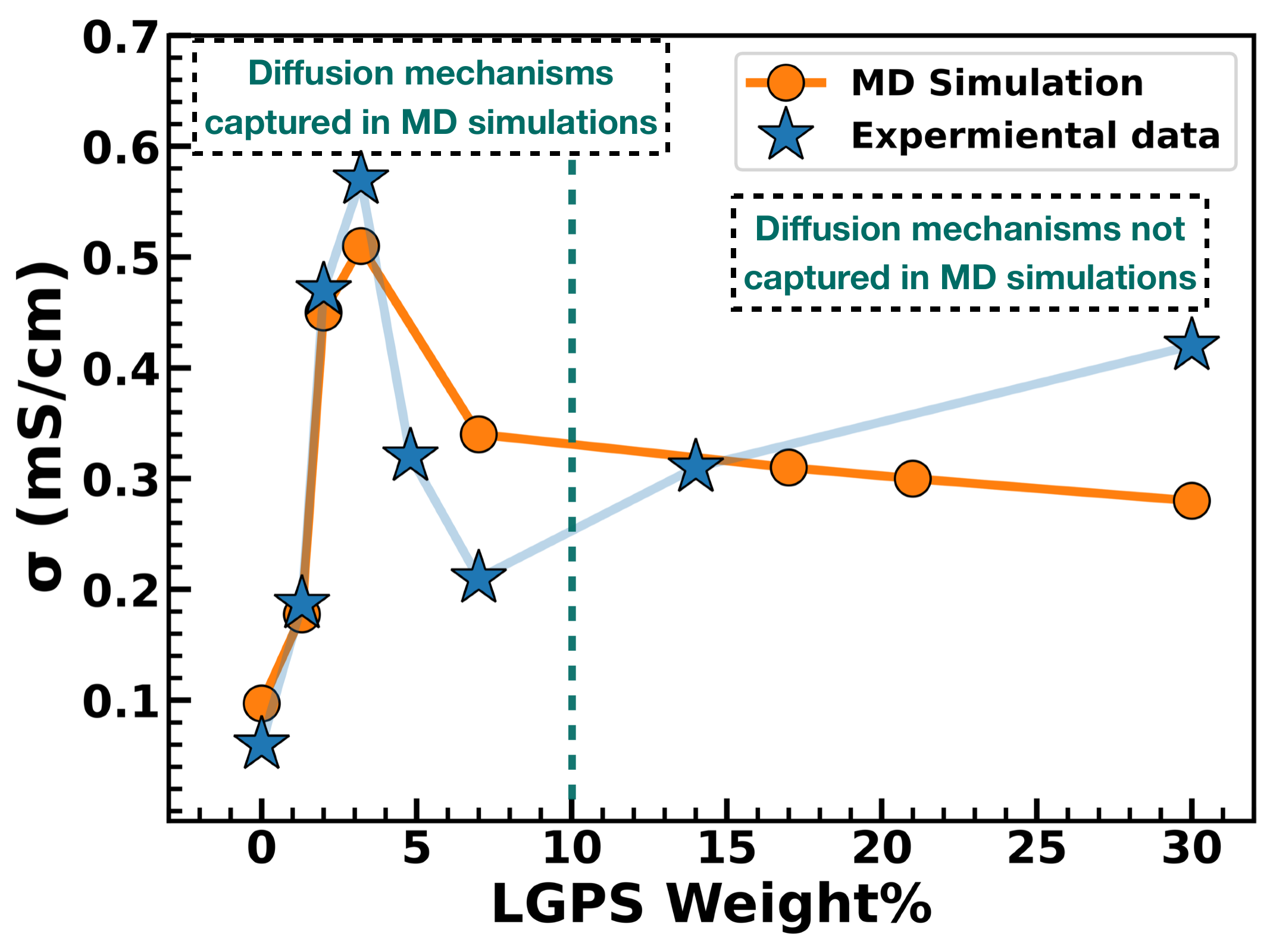}
    \caption{Ionic conductivity as a function of the LGPS weight ratio, with respect to the experimental composition presented in Table S2 and scaled for computational feasibility in Table 1, from MD simulations compared with experimental data (see Figure S1 for the experimental data shown in this plot as well as other data). Note that the result for 0 wt\% is from Ref. \cite{kondori2023room} and the 3.2\% results differs slightly from that in Ref. \cite{kondori2023room} due to uncertainties in measurements (see SI).}
    \label{fig:MD2}
\end{figure*}
The Green-Kubo (GK) approach calculates the real ionic conductivity $\sigma^{real}$ by taking the autocorrelation of the ionic current $\mathbf{J}$ in the electrolyte:
\begin{equation}
\label{eq:1}
\sigma^{real}= \frac{V}{k_{B} T} \int_{0}^{\infty} dt \langle \mathbf{J}(t) \cdot \mathbf{J}(0) \rangle
\end{equation}
\begin{equation}
\mathbf{J}(t) = q\sum_{i}^{N} z_{i}\mathbf{v}_{i}(t)
\end{equation}
Here, $q$ is the elementary charge, $z_i$ is the charge number (valence) of ion $i$, $\mathbf{v}_{i}$ is the velocity of ion $i$, $T$ is the temperature, $k_B$ is the Boltzmann constant, $V$ is the volume of the simulation box, and $N$ is the number of ions.
The Green-Kubo relations can also be expressed in terms of particle positions rather than velocities. This form, analogous to computing self-diffusion coefficients from the mean-squared displacement of particle positions, is used to compute $L^{ij}$ in this work \cite{lemaalem2023effects}:
\begin{equation}
L^{ij} = \frac{q^{2}}{6k_{\mathrm{B}}TV}\lim_{t\to\infty}\frac{d}{dt} \big< \sum_{\alpha}[\boldsymbol{r}_i^{\alpha}(t)-\boldsymbol{r}_i^{\alpha}(0)]\cdot \sum_{\beta}[\boldsymbol{r}_j^{\beta}(t)-\boldsymbol{r}_j^{\beta}(0)]\big>
\end{equation}
where $k_{\mathrm{B}}T$ is the thermal energy and $\boldsymbol{r}_i^{\alpha}$ is the position of particle $\alpha$ of type $i$ relative to the system's center of mass.
We also compute the self and distinct components of the diagonal transport coefficients $L^{ii}$. The self component is computed via:
\begin{equation}
L^{ii}{\mathrm{self}} = \frac{q^{2}}{6k_{\mathrm{B}}TV}\lim_{t\to\infty}\frac{d}{dt} \sum_{\alpha}\big< [\boldsymbol{r}_i^{\alpha}(t)-\boldsymbol{r}_i^{\alpha}(0)]^2\big>
\end{equation}
The distinct component can be computed by $L^{ii}_{\mathrm{distinct}} = L^{ii} - L^{ii}_{\mathrm{self}}$. The self terms are related to the self-diffusion coefficients $D_i$ via $L_{\mathrm{self}}{}^{ii} = \frac{D_i c_i}{k_{\mathrm{B}}T}$.
Assuming both the cation and the anion are univalent, $\sigma^{real}$ can be expressed as:
\begin{equation}
\sigma^{real}=L^{++}+L^{--}-2L^{+-}
\end{equation}
\begin{figure*}[!ht]
    \centering
    \includegraphics[width=1\textwidth,]{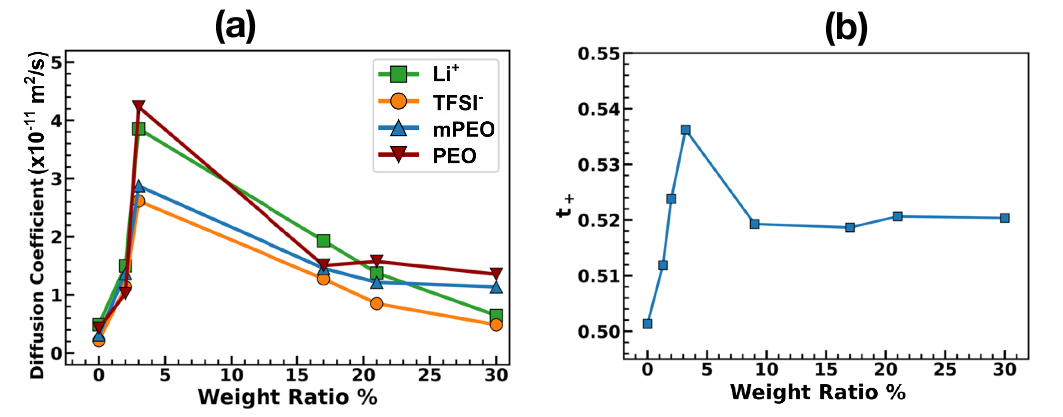}
    \caption{(a) Diffusion coefficients of CPE components and (b) Li$^{+}$ transference number as a function of LGPS weight ratio from MD simulations.}
    \label{fig:MD3}
\end{figure*}
The Nernst-Einstein ionic conductivity can be expressed using $L^{ii}_{\mathrm{self}}$ as:
\begin{equation}
\sigma^{NE}=L_{\mathrm{self}}{}^{++}+L_{\mathrm{self}}{}^{--}
\end{equation}
The cation transference number is assessed by:
\begin{equation}
t_{+}= \frac{D_{\mathrm{Li}^+}}{D_{\mathrm{Li}^+}+D_{\mathrm{TFSI}^-}}
\end{equation}
The molecular dynamics (MD) simulations for varying LGPS nanoparticle concentrations reveal a complex dependence of ionic conductivity on LGPS nanoparticle loading (see Figure~\ref{fig:MD2}). Experimental data for a similar system based on LGPS particles, LiTFSI, mPEO-TMS, and PEO is shown in Figure 3. In this figure, the measured ionic conductivities are plotted as a function of LGPS loading for comparison with the calculated values. Details of the experimental measurements are given in the SI. 
For LGPS weight fractions up to 10 x\%, the ionic conductivities calculated from MD simulations using the Green-Kubo relation agree with the measured values shown in (Figure~\ref{fig:MD2}), demonstrating a sharp fivefold increase in conductivity between 0 and 3.2 x\%  LGPS, followed by a decline past this optimal loading (Figure~\ref{fig:MD2}). This behavior reflects classical conduction mechanisms, enhanced segmental mobility, nanoconfinement effects, and favorable polymer/nanoparticle interfaces, that promote ion transport and are well-captured by MD. For computational feasibility, the LGPS particle size was rescaled to 1.2 nm. We acknowledge that this results in a higher surface-to-volume ratio compared to experimental crystallites ($\approx$ 17 nm), which may quantitatively amplify interfacial ion dynamics and the resulting cation transference number ($t_+$). However, this scaling preserves the qualitative physics of polymer segmental dynamics and interfacial saturation essential for understanding the low-loading regime. Polymer chain and ionic diffusion coefficients are highly correlated (Figure~\ref{fig:MD3}(a)), suggesting a global enhancement of polymer dynamics due to LGPS addition. We note that the excess mPEO-TMS, which exceeds the amount required for surface passivation, acts as an active component of the bulk electrolyte. Its structural similarity to PEO ensures miscibility, contributing to the bulk conductivity in the polymer-rich regime. The potential effect of mPEO-TMS inclusion on making the LGPS surface more amenable to ion transport is shown in the DFT calculations reported in Section 2.3. The MD simulations show that Li$^+$ ions diffuse more rapidly than TFSI$^-$ (Figure~\ref{fig:MD3}(a) and (b)), indicating the selective formation of conduction pathways along polymer and nanoparticle interfaces. These results are consistent with the literature: LGPS nanoparticles improve conductivity via interfacial interactions that increase free volume, reduce crystallinity, and optimize Lewis acid–base effects \cite{fu2022ion, sand2024critical, zhenginterface}. 
\begin{figure*}[!ht]
    \centering
   \includegraphics[width=0.6\textwidth,]{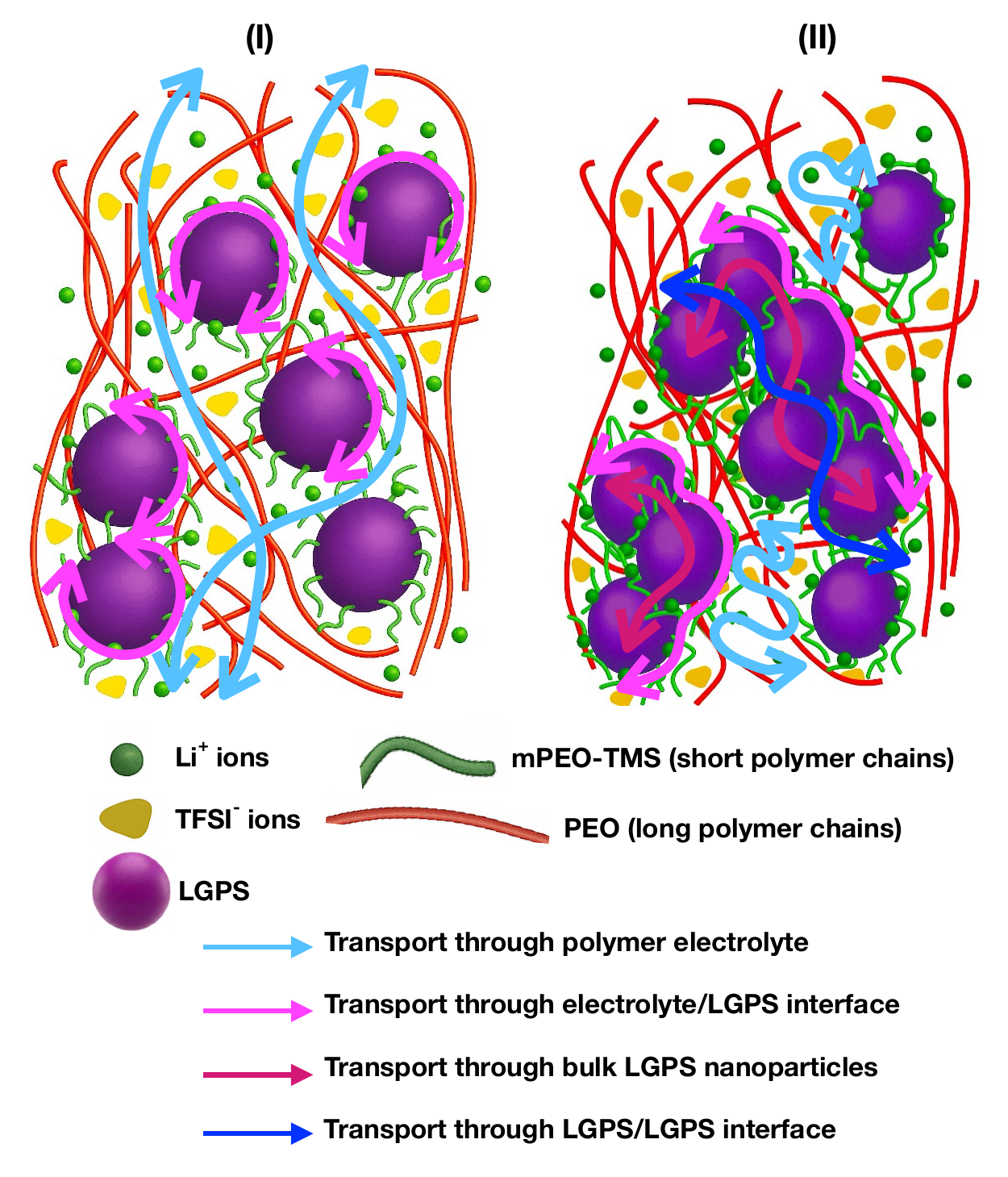}
    \caption{Schematic illustration of Li-ion diffusion and transport pathways in PEO/mPEO‑TMS–LGPS composite solid polymer electrolytes: (I) sparse LGPS network and (II) percolated LGPS-rich network.}
    \label{fig:Shema}
\end{figure*}
However, excessive nanoparticle content leads to agglomeration, disrupting continuous ionic pathways and decreasing conductivity. The optimal loading maximizes the interfacial area while maintaining network connectivity. At higher LGPS concentrations (x =~20\% or greater), experimental data show an enhancement in conductivity, whereas MD simulations do not show the increase. This discrepancy points to additional ion transport mechanisms that are not accessible to classical MD, particularly effects tied to concerted Li transport through ionically conducting LGPS agglomerates or on their interfaces. \\

The two ionic conduction regimes deduced from the MD simulations (Figure~\ref{fig:MD1}) and from the experiments (Figure~\ref{fig:MD2}) are illustrated in Figure~\ref{fig:Shema}. In regime I, the conduction pathways are through the polymers and the polymer/LGPS interfaces, giving rise to a volcano type curve as a function of LGPS wt\%. The MD simulations suggest that the decrease in ionic conductivity above 3.2 wt\% results from the formation of LGPS clusters that interrupt polymer-based pathways, whereas the increase below 3.2 wt\% reflects the enhanced contribution of polymer/interface transport and segmental diffusion \cite{rajahmundry2024understanding, choo2020diffusion}, which is a polymer-type diffusion process associated with the local motion of polymer chain segments (Figure~\ref{fig:MD3}). At LGPS contents greater than 20 wt\% the experimental results show that the ionic conductivity starts increasing again. The discrepancy between MD simulations and experimental results at loadings above 20\% wt highlights the limitations of classical molecular dynamics. The classical force fields (UFF/OPLS-AA) treat LGPS particles as high-impedance materials because they lack appropriate parameterization for ion diffusion through the LGPS bulk and for the vacancy-mediated Li hopping mechanism at the LGPS surface. Consequently, classical MD fails to capture the experimentally observed recovery in ionic conductivity at high loadings ( $\ge$ 20\% wt), where transport becomes dominated by the inorganic phase. This regime involves Li-ion conduction through percolated ceramic networks, consistent with the spectral signatures reported by Li \textit{et al.} \cite{Li2021}. We note that the present DFT calculations describe only the surface diffusion mechanism and do not account for bulk ion transport.\\

\subsection{Structural properties from MD simulation}
Structural analysis employed the Radial Distribution Function (RDF), g(r), and the Coordination Number N(r) to examine the spatial particle distribution and local structural organization in the electrolyte.
The RDF is defined as:
\begin{equation}
g_{\alpha \beta}(r)=\frac{\langle \rho_\beta(r) \rangle}{\langle\rho_\beta\rangle_{local}}=\frac{1}{\langle\rho_\beta\rangle_{local}}\frac{1}{N_{\alpha}}
\sum_{i \in \alpha}^{N_{\alpha}} \sum_{j \in \beta}^{N_{\beta}}
\frac{\delta( r_{ij} - r )}{4 \pi r^2}
\end{equation}
Where $\langle\rho_B(r)\rangle$ is the particle density of type $\beta$ at a distance r from particles $\alpha$, and $\langle\rho_B\rangle_{local}$ is the average particle density of type $\beta$ within a radius $r_{max}$ (12\AA{}).\\
The coordination number as a function of distance, N(r), is expressed as:
\begin{equation}
N(r) = 2\pi n_b \int_0^r g(r') r' dr'
\end{equation}
where r represents an arbitrary distance from a reference particle, and $n_b$ represents the bulk density.\\

\begin{figure*}[!ht]
    \centering
    \includegraphics[width=1\textwidth,]{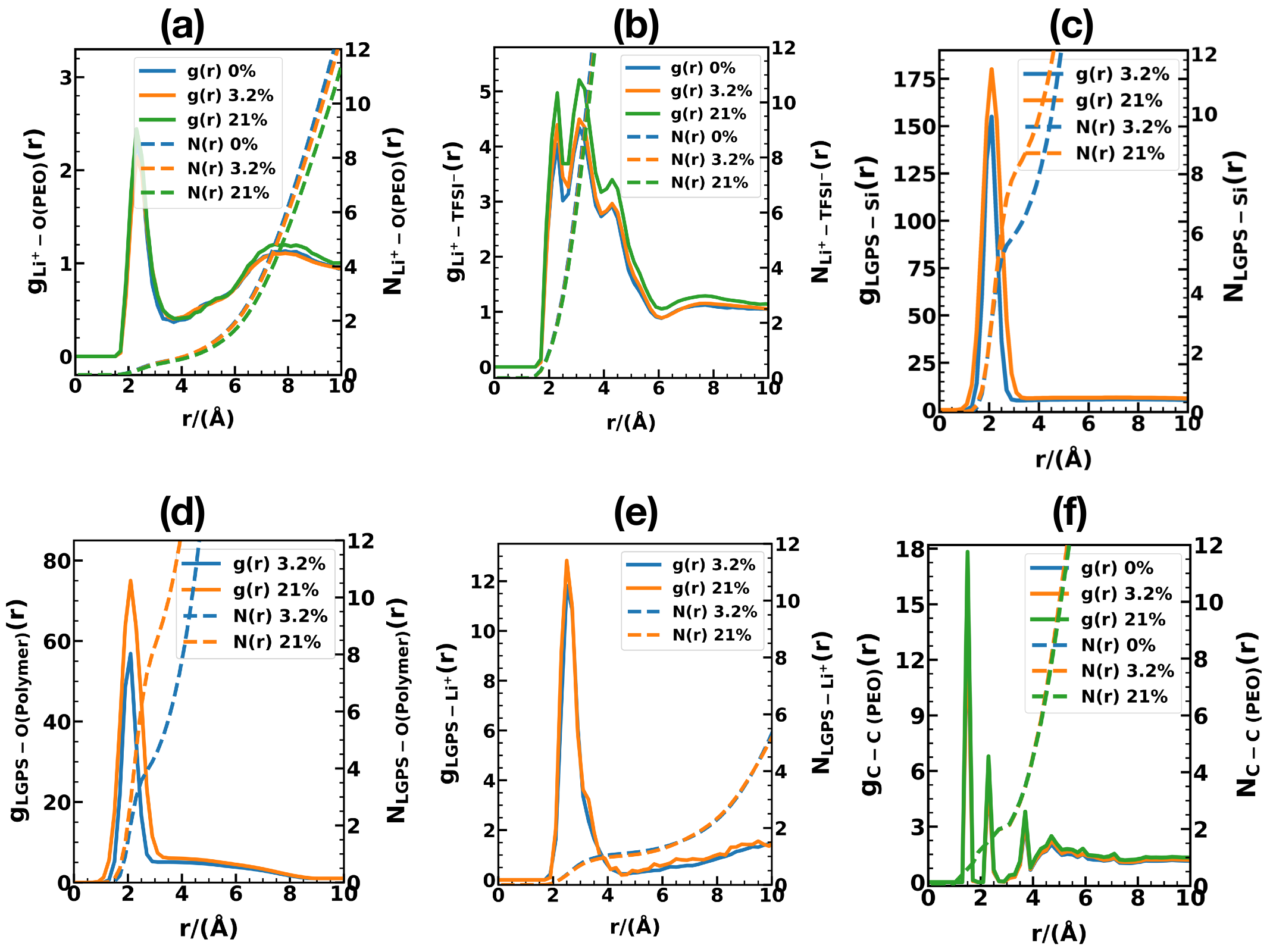}
    \caption{(a) Partial radial distribution functions (RDF), g(r), and running coordination number N(r) showing moderate interaction between Li$^+$ and PEO. (b) RDF and N(r) indicating a moderate interaction between Li$^+$ and TFSI$^-$. (c) RDF and N(r) revealing strong interactions between PEO and LGPS via the interactions between LGPS nanoparticles and Si atoms. (d) RDF and N(r) revealing strong interactions between PEO, mPEO-TMS, and LGPS mediated by O atoms from the polymers. (e) RDF and N(r) indicating strong interaction between LGPS nanoparticles and Li$^+$ ions. (f) RDF and N(r) for the carbon atoms in poly(ethylene oxide) (PEO) (Figure S2 depict RDF and N(r) for the other pairs). The RDF and N(r) are presented at various LGPS loading (0 x\%, 3.2 x\%, and 21 x\%), illustrating local structural environments and microstructural changes.}
    \label{fig:MD4}
\end{figure*}

Figure~\ref{fig:MD4} (a)-(e) presents a comparative analysis of the radial distribution functions (RDFs) and coordination numbers \( N(r) \) for key atom pairs in polymer electrolytes containing different LGPS nanoparticle contents (0 x\%, 3.2 x\%, and 21 x\%). The RDF between Li\(^+\) ions and PEO oxygen atoms (Figure~\ref{fig:MD4}(a)) shows a distinct primary peak at \( r \approx 2.3 \, \text{\AA} \) for all systems, indicating moderate and consistent Li\(^+\)–PEO interactions. The nearly unchanged peak height and coordination number suggest that increasing LGPS content up to 21 x\% does not significantly alter Li\(^+\) solvation by PEO, preserving the solvation shell structure. The Li\(^+\)–O(TFSI\(^-\)) RDF (Figure~\ref{fig:MD4}(b)) exhibits a dominant peak at \( r \approx 2.4 \, \text{\AA} \) across all compositions, with slightly higher intensity at 21 x\% LGPS, implying enhanced Li\(^+\)–anion association at higher nanoparticle loading. Figure~\ref{fig:MD4}(c and d) shows that at 21 x\% LGPS, the RDF between Si(mPEO-TMS) and Li, as well as O(mPEO) and Li atoms at the LGPS surface, displays a sharp and intense peak, indicating strong and specific (mPEO-TMS)–LGPS interactions driven by polymer adsorption and interfacial structuring. This interface adsorption will be further studied using DFT calculations for bond formation possibilities in the next section. At 3.2 x\% LGPS, the peak is less intense, consistent with less organized contacts between PEO and LGPS compared to 21 x\%. The more intense sharp peak at 21 x\% signifies a higher contact of PEO segments with nanoparticle surfaces, likely due to the increased interfacial area and possible particle aggregation. Figure~\ref{fig:MD4}(e) presents the Li$^+$-NP (LGPS nanoparticle) radial distribution functions, where the observed peaks for both concentrations are closely aligned. This indicates that the local structural environment of Li$^+$ near the nanoparticles remains highly similar between the two LGPS concentrations.

Figure~\ref{fig:MD4} (f) and Figure S2 describe how specific atomic pairs, such as C–C, H–H, C–O, and O–O, are spatially distributed within the polymer as a function of distance. Each plot in the figure shows both the RDF, $g(r)$, and the corresponding running coordination number, $N(r)$, for these atomic pairs under varying additive concentrations (0 x\%, 3.2 x\%, and 21 x\%), reflecting microstructural changes.
Sharp initial peaks in the C–C and C–O RDFs, seen in Figure~\ref{fig:MD4}(f) and Figure S2, originate from neighboring backbone atoms, with characteristic short-range C–C peaks at $r \approx$ 1.5 \text{\AA}, while the pronounced features in the H–H and O–O RDFs presented in Figure S2 reflect local hydrogen and oxygen environments. As the additive content increases, subtle but visible changes appear, such as increased peak heights, especially at the highest concentration (21 x\%), indicating a more congested polymer. The running coordination number curves, however, show no significant variation across the studied concentrations.
The Li\(^+\) cluster size within the electrolyte varies non-monotonically with LGPS content (x\%) (Figure S3) due to the competing effects of polymer flexibility and nanoparticle dispersion. At low x\% (<3.2), the addition of LGPS increases the polymer diffusion coefficient, boosts ionic conductivity, and leads to a more uniform Li$^+$ distribution with slightly smaller clusters. At higher x\% (>3.2), nanoparticle agglomeration impedes conduction pathways and restricts polymer motion, promoting Li$^+$ clustering near LGPS agglomerates or LGPS/Polymer interfaces and increasing the average cluster size.

\subsection{mPEO-TMS|LGPS interface properties from DFT calculations}
\begin{figure*}[!ht]
    \centering
    \includegraphics[width=0.7\textwidth,]{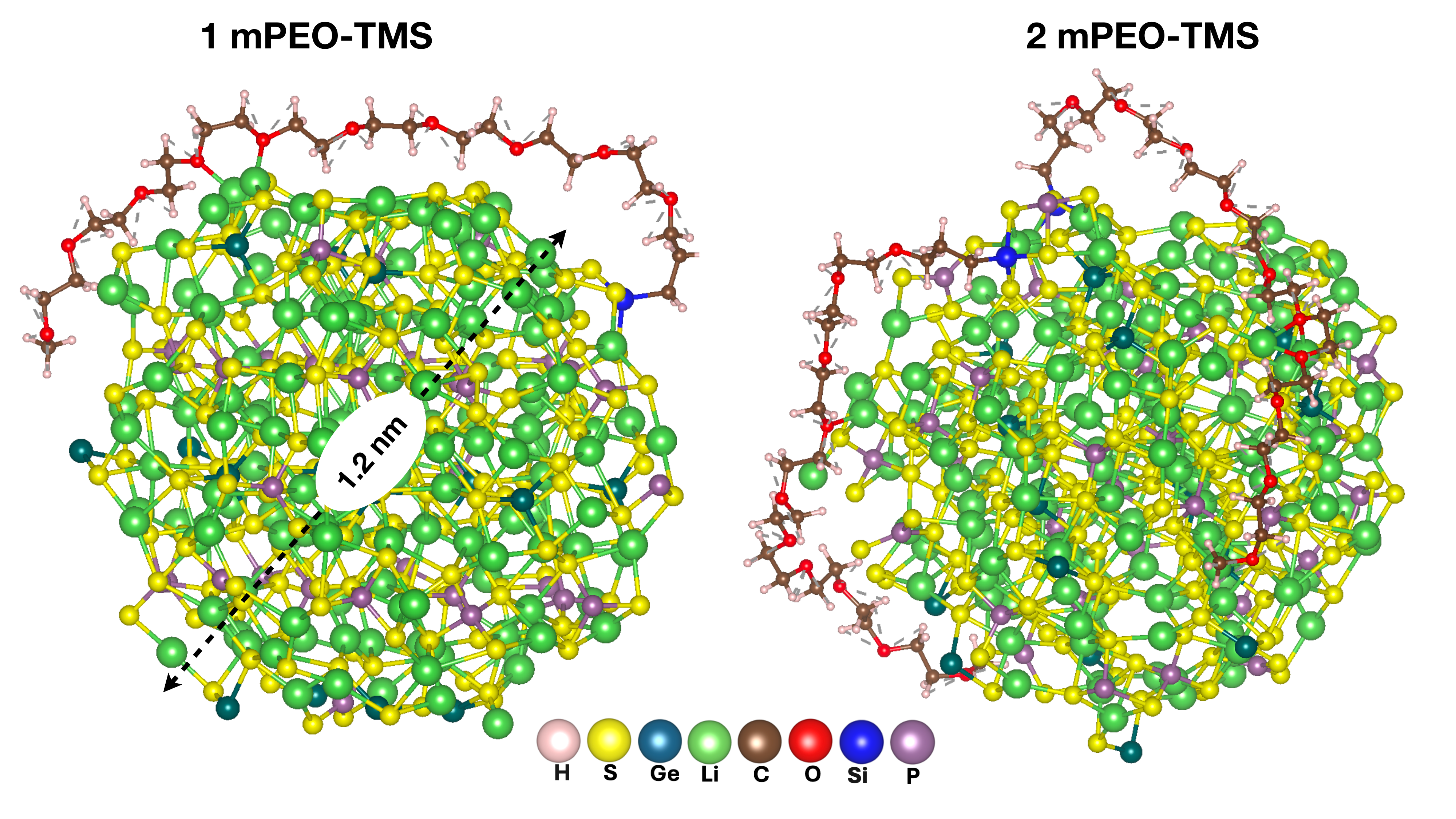}
    \caption{Optimized structures showing a) 1 molecule of mPEO-TMS  b) 2 molecules of mPEO-TMS wrapped around the LGPS nanoparticle.}
    \label{fig:FigDFT2}
\end{figure*}

The objective of our DFT calculations was to determine whether the mPEO-TMS|LGPS system creates Li-ion transport channels with low barriers to Li-ion transport, which is one of the keys to favorable polymer/nanoparticle interfaces suggested by the MD simulations. We first investigated how the mPEO-TMS chain interacts with LGPS beyond the surface SiS$_3$ bond. A nanoparticle structure with one mPEO-TMS molecule was optimized, and we observed that the molecule follows the surface of LGPS, wrapping around the nanoparticle. Some Li-O bonds are formed along the interface, enhancing the interaction between the nanoparticle and mPEO-TMS. Figure~\ref{fig:FigDFT2} shows that adding a second mPEO-TMS molecule exhibits similar effects. 
We focused our investigations on determining the ability of Li ions on this surface to move along the Li–O bond chains formed. We propose a Li vacancy-driven mechanism, whereby the movement of the ions is influenced by adjacent vacancies. The surface Li migration mechanism investigated here assumes the presence of Li ion vacancies on the surface.  We expect that the LGPS surface will have Li vacancies with little energy cost just as is present in LGPS bulk that is responsible for its high bulk ionic conductivity \cite{wang2015design}. In addition, the three Li released while grafting the mPEO-TMS units to the LGPS particles will also contribute to the presence of vacancies present on the surface. The barrier for the Li ion to move from its initial position to a newly formed vacancy was determined using the NEB method described in the computational details, and only a single mPEO-TMS chain was included in the Li migration calculations. We observed that the chemical environment on the surface of LGPS plays a key role in determining the minimum energy paths. The first structure investigated showed that it is feasible for Li to move from one oxygen atom to the next with a Li vacancy along the mPEO-TMS chain, provided that sufficient S atoms are present to form either a LiO–2S or LiO–3S bond. When atoms such as Ge are present along the path, the energy barrier is high since Li fails to form sufficient bonds with the LGPS surface.
\begin{figure*}[!ht]
    \centering
    \includegraphics[width=1\textwidth,]{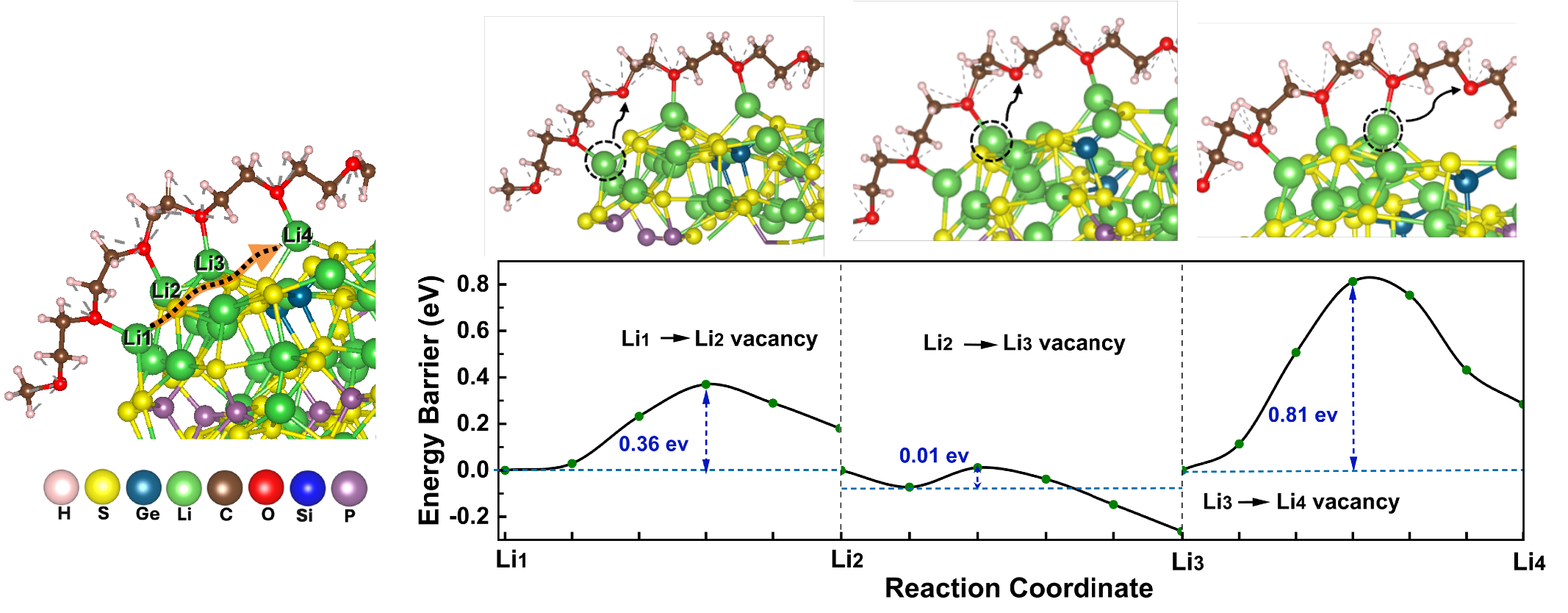}
    \caption{Vacancy-driven Li cation migration via sequential non-concerted hops and the corresponding energy barrier.}
    \label{fig:FigDFT3}
\end{figure*}
The path followed for the mPEO-TMS|LGPS structure is shown in Figure~\ref{fig:FigDFT3}, along with the corresponding energy barriers. Figure~\ref{fig:FigDFT3} indicates that the energy barrier is low for the Li1–Li2 (\(\sim 0.37\,\text{eV}\)) and Li2–Li3 (\(\sim 0.08\,\text{eV}\)) paths, but significantly larger for the Li3–Li4 (\(\sim 0.81\,\text{eV}\)) path. Li1–Li2 and Li2–Li3 have similar environments, with no other non-Li cation nearby. In contrast, Li3–Li4 has Ge in the vicinity of the proposed path and insufficient sulfur for Li to form bonds while moving to the new position. We altered the surface environment to determine whether changes occurred in the migration barrier along this path. Figure~\ref{fig:FigDFT4} shows that the energy barrier, as the Li cation hops from one vacancy to another, remains below \(0.5\,\text{eV}\). The difference between the structures in Figure~\ref{fig:FigDFT3} and~\ref{fig:FigDFT4} is the atomic composition at the mPEO-TMS|LGPS interface. Figure~\ref{fig:FigDFT4} shows that more S atoms are present on the surface and are available to bond with the Li cation as it moves along the path. This effect is especially seen in the Li4–Li5 vacancy migration, where an S atom shields the Li ion from the effect of Ge along the path, a phenomenon observed in Figure~\ref{fig:FigDFT3} for the Li3–Li4 migration.
\begin{figure*}[!ht]
    \centering
    \includegraphics[width=1\textwidth,]{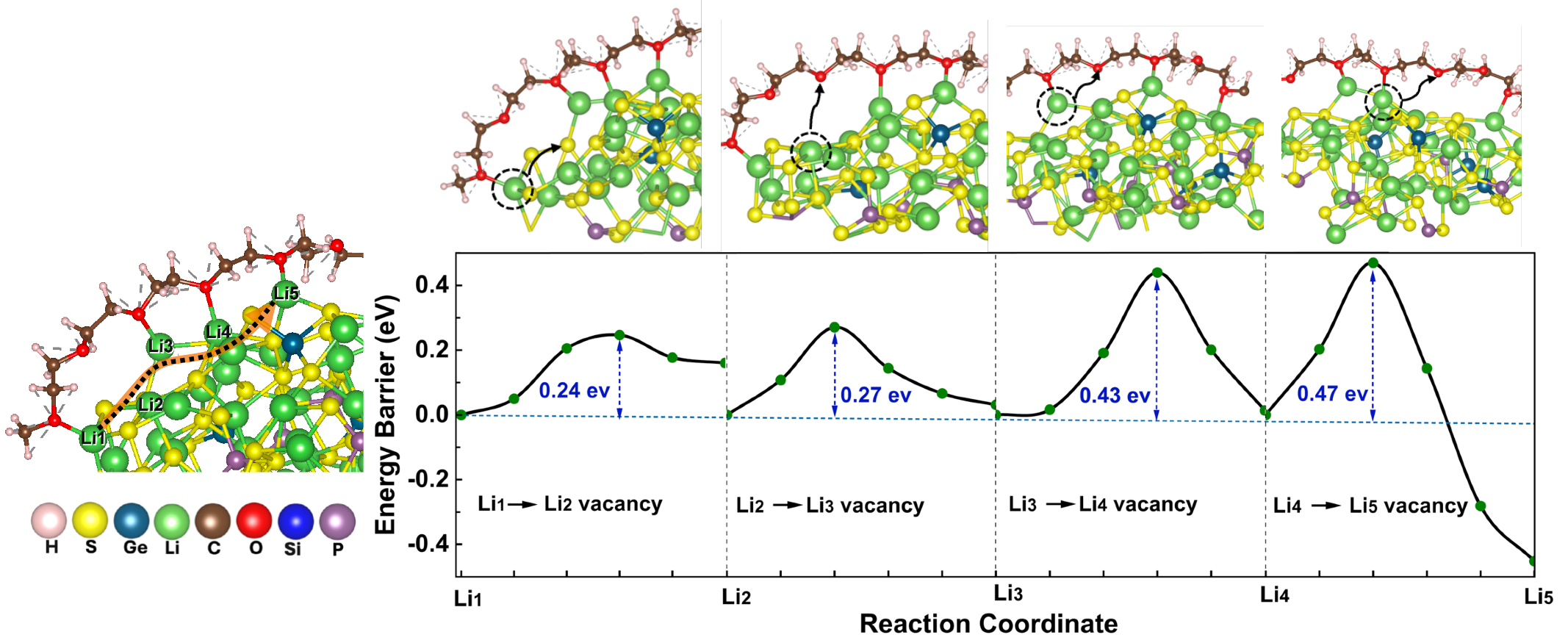}
    \caption{The effect of modifying the chemical environment of the mPEO-TMS|LGSP interface on Li-ion transport.}
    \label{fig:FigDFT4}
\end{figure*}
Another surface modification investigation is outlined in the supplementary information (Figure S4). In this structure, Li1 is bonded to two oxygen atoms. The Li–S bonds on the surface are modified, and no Ge is in the vicinity of the proposed path. We observed that the energy barrier for Li1–Li2 is approximately \(0.39\,\text{eV}\), comparable to the Li1-Li2 energy barrier value ($\approx$ \(0.37\,\text{eV}\)) shown in Figure~\ref{fig:FigDFT3}, although the energy profile plateaus early. Li2–Li3 shows a barrier of \(\sim 0.11\,\text{eV}\), while Li3–Li4 decreases from \(\sim 0.8\,\text{eV}\) (Figure~\ref{fig:FigDFT3}) to \(\sim 0.26\,\text{eV}\), which is attributed to the absence  of Ge along the migration pathway. This is because the Ge-S environment causes Li to interact more strongly with the polyethylene oxide oxygens to inhibit migration of the Li ions.  \\

Our DFT calculations, therefore, indicate that Li-ion transport channels can exist at the mPEO-TMS|LGPS interface; however, they are not continuous. Li ions can move to adjacent vacancies if the chemical environment keeps the cation sufficiently bonded along the path. Thus, our DFT calculations indicate that Li-ion transport channels can have low barriers to Li-ion transport and contribute to ionic conductivity pathways. The low-barrier pathways we identified are primarily associated with sulfur-rich pathways, which are statistically abundant given that there are many more sulfur atoms in the LGPS crystal structure. In a realistic percolation network, these sites are not isolated but form quasi-continuous channels. Even if interrupted by occasional high-barrier (Ge-rich) sites, the percolating nature of the network should allow Li ions to detour through the abundant low-barrier S sites.
\section*{Conclusion}

This study elucidates how LGPS nanoparticle loading modulates Li-ion transport in composite solid polymer electrolytes. Molecular dynamics reveal a sharp increase in ionic conductivity up to 3.2 x\% of LGPS, followed by a decrease governed by classical polymer-segmental and interface-driven transport mechanisms. The volcano-type curve at low LGPS wt\% is in agreement with the experimental results. At higher loading levels (20\% or more of LGPS), experiment shows a renewed increase in ionic conductivity that MD simulations fail to capture, indicating a shift to a ceramic-dominated transport mechanism. DFT calculations elucidate that Li-ion migration proceeds via vacancy-driven hopping. This transport channel, distinct from the classical polymer dynamics, becomes dominant at high loadings where nanoparticles agglomerate to form percolated, bulk-like domains, driving the observed conductivity increase.

Optimal ionic transport in the LGPS-polymer electrolyte relies on engineered polymer-ceramic interfaces and effective nanoparticle dispersion, as excessive loading leads to agglomeration and hinders conductivity through the polymer electrolyte. DFT results demonstrate that specific chemical environments can lower the migration barrier, which can lead to more favorable transport channels consistent with findings from the MD simulations. The DFT calculations indicate that mPEO/LGPS  migration barriers  are dictated by the local sulfur and germanium environments. This offers new design pathways for the improved performance of composite polymer electrolytes.

\section*{Supporting Information}
Details on the Molecular Dynamics simulation setup, including force fields and equilibration protocols; tables summarizing molecular masses and experimental electrolyte compositions; and additional figures displaying experimental ionic conductivity, radial distribution functions (RDFs) for PEO, Li-ion cluster analysis, and illustrations of interface modifications.
\section*{Acknowledgements}This research is supported by the Vehicle Technologies Office (VTO), Department of Energy (DOE), USA, through the Battery Materials Research (BMR) program. Argonne National Laboratory is operated for DOE by UChicago Argonne, LLC under the contract number DE-AC02-06CH11357. The authors gratefully acknowledge the computing resources provided on Bebop and Improv, high-performance computing clusters, operated by the Laboratory Computing Resource Center at Argonne National Laboratory.

\bibliography{Refs}
\end{document}